\documentclass[%
 reprint,
showpacs,preprintnumbers,
 amsmath,amssymb,
 aps,
]{revtex4-1}

\usepackage{graphicx}
\usepackage{dcolumn}
\usepackage{bm}
\usepackage{hyperref}
\usepackage{slashed}
\usepackage{mathrsfs}
\usepackage{lineno}
\usepackage{enumitem}
\newcommand{\be}{\begin{equation}}
\newcommand{\ee}{\end{equation}} 
\newcommand{\bes}{\begin{equation*}}
\newcommand{\ees}{\end{equation*}}
\newcommand{\bra}{\langle}
\newcommand{\ket}{\rangle}

\begin{document}

\preprint{APS/123-QED}

\title{Strong, random field behavior of the Euclidean Dirac propagator}

\author{M. P. Fry}
\affiliation{%
University of Dublin, Trinity College, Dublin 2, Ireland\\
}%




\date{\today}

\begin{abstract}
The one-loop effective action of quantum electrodynamics in four dimensions is shown to be controlled by the Euclidean Dirac propagator $G$ in a background potential. After separating the photon self-energy and photon-photon scattering graphs from the effective action the remainder  is known to be the logarithm of an entire function of the electric charge of order 4 under mild regularity assumptions on the potential. This input together with QED's lack of an ultrastable vacuum constrain the strong field behavior of $G$. It is shown that $G$ vanishes in the strong field limit. The relevance of this result to the decoupling of QED from the remainder of the electroweak model for large amplitude variations of the Maxwell field is discussed.
\end{abstract}

\maketitle


It may seem surprising that anything more remains to be said about the quantized Euclidean Dirac field propagator $G$ in four dimensions in a background potential $A_{\mu}$, where
\be \label{G}
\begin{split}
&G(x,y) = [ -(\slashed{p} - e\slashed{A})+m] \\
&\times   \int_{0}^{\infty} dt ~ e^{-tm^2} \bra x| e^{-t[ (p-eA)^2 + e\sigma F/2 ]} | y \ket,
\end{split}
\ee
$\slashed{p} - e\slashed{A}$ is the Euclidean Dirac operator $\slashed{D}$ with anti-Hermitian $\gamma$-matrices, $\gamma_{\mu}^{\dagger} = -\gamma_{\mu}$, with $\{\gamma_{\mu}, \gamma_{\nu}\} = -2\delta_{\mu \nu}$ and $\sigma_{\mu \nu} = [\gamma_{\mu}, \gamma_{\nu}]/{(2i)}$. Excepting the general result (\ref{bound}) below there are still no results for the asymptotic behavior of $G$ for strong fields on $\mathbb{R}^4$, including random ones. Intuition is helped by viewing $F_{\mu \nu}$ in Euclidean space as a time-independent four-dimensional magnetic field. Competition between the diamagnetic $(p-eA)^2$ and paramagnetic $e\sigma F/2$ terms in the exponentiated Hamiltonian in (\ref{G}) remains an obstruction to the strong field analysis of $G$. It is the aim of this paper to demonstrate that $G$ vanishes in the strong field limit for a broad class of potentials. It will be explained below why this is of physical interest. 

The term strong field in this paper refers to the large amplitude variation of a random potential that occurs in a Euclidean functional integral over $A_{\mu}$. $G$'s strong-field behavior has no immediate connection with the Minkowski Dirac propagator. For example, the continuation of the four-dimensional magnetic field in $G$ to the Minkowski metric can result in imaginary time electric and magnetic fields. When the continuation results in physical fields then a subset of them may simulate laser pulses. Then $G$ becomes the propagator of a charged particle in such a background that can be relevant to current experiments with extremely high-intensity lasers \cite{00}. A theorem is needed that ensures the continued $G$ also vanishes in the strong-field limit. This would be a highly non-trivial result, considering how difficult it is to calculate the Euclidean $G$'s strong field limit as discussed in comment 4 below.

The most restrictive bound on $G$ known to the author is that of Vafa and Witten \cite{VW} :
\be \label{bound}
|\bra \alpha |G|\beta \ket| \leq \frac{e^{-mR}}{m} \sqrt{\bra \alpha | \alpha \ket} \sqrt{\bra \beta | \beta \ket}.
\ee
Here $| \alpha \ket$ and $| \beta \ket$ are any two states of disjoint support, separated by a minimum distance $R$. The bound is remarkable for its generality: $A_{\mu}$ can be a random potential with no particular symmetry subject to the constraint that it is regular enough for $i\slashed{D}$ to be self-adjoint on a suitable function space. It establishes that $G$ cannot have unbounded growth in a strong magnetic field.

The strong-field asymptotic behavior of $G$ is relevant to the extraction of non-perturbative information from the electroweak model. This renormalizable model with its $17+7$ adjustable parameters, including three massive Dirac neutrinos and their mixing, is a complete theory of the electroweak interaction. Perturbation expansions are reliable as long as the coupling constants remain small. Do they? Is there a Landau pole \cite{000}? How does the renormalization group equation for the Higgs coupling behave when summed \cite{0000}? These are some of the questions perturbation theory cannot answer. The reader who wishes to avoid further discussion of the electroweak model may proceed to the paragraph beginning above Eq.(9).


Whether it can be non-perturbatively quantized depends on the convergence of the unexpanded Euclidean functional integrals over all classical field configurations for the vacuum expectation value of its field operators. Integrating out the fermion degrees of freedom results in an effective action depending on 6 lepton and $3\times 6$ quark determinants from the neutral weak-current and $3+3\times3$ determinants from the charged weak-current, including color, that are functionals of the Higgs, Maxwell, $Z$ and $W$ fields \cite{Fry}. Sense can be made of these ill-defined determinants by regularization, renormalization, and the cancellation of their embedded chiral anomalies. When this is done there remain the functional integrals over the Higgs and gauge fields. The question has been asked whether any of these integrals converge \cite{Fry}. Assuming that the order of doing the integrals is arbitrary it was decided to integrate over the Maxwell field first. Convergence hinges on the growth of the QED one-loop effective action for large amplitude variations of $F_{\mu \nu}$ and hence $A_{\mu}$ provided QED decouples from the remainder of the electroweak model in this limit \cite{Fry}. Faster than quadratic growth in $F_{\mu \nu}$ would place in doubt whether any process can be calculated non-perturbatively in the electroweak model that includes dynamical fermions. Decoupling happens when the remainder of the electroweak model's effective action grows no faster than quadratically with $F_{\mu \nu}$.

To see how decoupling can occur consider the determinants contributed by the neutral weak-current
 \begin{widetext}
 \be \label{3}
 \begin{split}
&\mathrm{ln\,det}\left[  \slashed{p}+m_i - Q_ie \slashed{A} - \frac{g}{2\cos \theta_W}(g_V - g_A \gamma_5)\slashed{Z} + \frac{gm_i}{2M_W} H \right] - \mathrm{ln\,det}(\slashed{p}+m_i) \\
=~& \mathrm{ln\,det} (1-Q_i eS \slashed{A}) + \mathrm{ln\,det}\left[ 1 + G\left( -\frac{g}{2\cos \theta_W}(g_V - g_A\gamma_5)\slashed{Z}  + \frac{gm_i}{2M_W} H \right) \right],
\end{split}
\ee
\end{widetext}
where $G = (\slashed{p} - Q_i e \slashed{A} + m_i)^{-1}$, $S = (\slashed{p}+m_i)^{-1}$, $g_V^i = t_{3L}(i) - 2Q_i \sin^2\theta_W$, $g_A^i = t_{3L}(i)$, $t_{3L}(i)$ is the weak isospin of fermion $i$, and $Q_i$ is its charge in units of the positron electric charge, $e$; $\theta_W$ is the weak angle; $g= e/ \sin \theta_W$; $m_i$ and $M_W$ are the fermion and $W$ mass, respectively. The conventions and notation of \cite{CPatrignani} are followed here.

The determinants contributed by the charged weak-current are, for quarks, 
\be \label{4}
\begin{split}
& \mathrm{ln\,det}\Big[ 1 - \frac{g^2}{8} G_{t_{3L}{(i)} = -1/2} \slashed{W}^{-} \\ 
& \times (1-\gamma_5)  G_{t_{3L}(i)=1/2 } \slashed{W}^{+}(1-\gamma_5) \Big],
\end{split}
\ee

and for leptons 
\be \label{5}
\begin{split}
&\mathrm{ln\,det}\Big[ 1 - \frac{g^2}{8} G_{t_{3L}{(i)} = 1/2} \slashed{W}^{+} \\ 
&\times (1-\gamma_5)  G_{t_{3L}(i)=-1/2 } \slashed{W}^{-}(1-\gamma_5) \Big],
\end{split}
\ee
where 
\be \label{6}
\begin{split}
&G_{t_{3L}(i) } = G \\
&~-G\left( -\frac{ g }{ 2\cos \theta_W } (g_V^i - g_A^i \gamma_5) \slashed{Z} + \frac{gm_i H}{2M_W}\right) G_{ t_{3L}(i) } .
\end{split}
\ee

Each determinant in (\ref{4}) and (\ref{5}) is for a quark pair or lepton pair belonging to the same family such as $(u,d)$, $(\nu_e, e)$, etc. Note the all-pervasive presence of $G$ in (\ref{3})-(\ref{6}). This has its origin in the factorization of the QED effective action from the electroweak model's that occurs in (\ref{3}).  It is assumed that quark mixing and neutrino mixing are irrelevant to the behavior of the effective action for the large amplitude variations of the Maxwell field.  

The first term on the right-hand side of (\ref{3}) contributes to the QED effective action considered below. The second term in (\ref{3}) and the determinants in  (\ref{4}) and  (\ref{5})  must be renormalized and their triangle graphs' chiral anomalies cancelled by summation over fermion families, including color. An example of the cancellation of an anomaly in a triangle graph is given in \cite{Fry} for $\gamma \to W^+W^-$, including quark mixing. Potential chiral anomalies from box graphs such as $AAA\gamma_5Z$ and $A\gamma_5 Z \gamma_5 Z \gamma_5 Z$ are removed by Euclidean $C$-invariance. By this we mean there exists a matrix $C$ such that $C \gamma_{\mu}C^{-1} = \gamma_{\mu}^T$. In the representation of the $\gamma$-matrices used in \cite{Fry}, Eq. (D7), $C = \gamma_3 \gamma_1$. These operations are done by expanding in $e$ and $g$ through fourth order giving a residue of terms of not more than $O((eF)^2)$.  The remaining terms are ultraviolet finite and $G$-dependent, and so a necessary condition for decoupling is that $G$ vanish for large amplitude variations of $A_{\mu}$. Therefore, information beyond the Vafa-Witten result (\ref{bound}) is required and is the aim of this paper.    

Eq. (9) and the result (15) below indicate that $G$ also controls QED's one-loop effective action. There are $3\times 2 + 3$ $G$s corresponding to the three families of quarks and charged leptons, neglecting color. We now turn to the effective action of QED and its relation to $G$.

As the potentials support a gauge-fixed Gaussian measure $\mu(A)$ on $\mathscr{S}'(\mathbb{R}^4)$, the space of tempered distributions, they are neither differentiable nor locally square-integrable. They will be smoothed by convoluting them with functions $f_{\Lambda}$ belonging to $\mathscr{S}(\mathbb{R}^4)$, the space of functions of rapid decrease:
\be
A_{\mu}^{\Lambda}(x) = \int {\rm d}^4y~ f_{\Lambda}(x-y)A_{\mu}(y).
\ee 
Then $A_{\mu}^{\Lambda} \in C^{\infty}$ and hence is infinitely differentiable. This smoothing process has the beneficial effect of introducing a gauge invariance preserving ultraviolet cutoff required to regulate QED. Thus, from the covariance of $\mu(A)$, $\int {\rm d}\mu(A)~ A_{\mu}(x) A_{\nu}(y) = D_{\mu \nu}(x-y)$, where  $D_{\mu \nu}(x-y)$ is the free photon propagator in a fixed gauge, obtain
\be
\int {\rm d}\mu(A)~ A_{\mu}^{\Lambda}(x) A_{\nu}^{\Lambda}(y) = D_{\mu \nu}^{\Lambda}(x-y),
\ee
where the regularizing photon propagator $ D_{\mu \nu}^{\Lambda}$ has the Fourier transform $ \hat{D}_{\mu \nu}(k) | \hat{f}_{\Lambda}(k)|^2$ with $\hat{f}_{\Lambda} \in C_0^{\infty}$, the space of $C^{\infty}$ functions with compact support such as $\hat{f}_{\Lambda}(k)=1$, $k^2 \leq \Lambda^2$ and $\hat{f}_{\Lambda}(k)=0$, $k^2 \geq n \Lambda^2$, $n >1$ \cite{Fry}. The $A_{\mu}^{\Lambda}$ replace $A_{\mu}$ everywhere in the functional integrals over $A_{\mu}$ except in the measure $\mu(A)$. In the following the superscript $\Lambda$ will be omitted with the understanding that $A_{\mu}$ is now a $C^{\infty}$ function. Only when it encounters the measure does $\Lambda$ reappear. We will deal with the falloff at infinity of the potentials supported by $\mu(A)$ below.  

Consider any of QED's renormalized fermion determinants contributing to its effective action corresponding to a specific quark or charged lepton. They can be defined as \cite{4}, \cite{5}, \cite{6}
\be \label{9}
\mathrm{ln\,det}_{\text{ren.}} = \Pi_2 + \Pi_4 + \mathrm{ln\,det}_5(1-eS\slashed{A}),
\ee
where $S$ is the free fermion propagator, $e$ is its electric charge, and $\Pi_2$ and $\Pi_4$ contain the renormalized photon self-energy and $\gamma \gamma$-scattering graphs, respectively. The determinant $\mathrm{det}_5$ is defined by \cite{7}, \cite{8}, \cite{9}, \cite{10}
\be \label{10}
\mathrm{ln\,det}_5(1-eS\slashed{A}) = \mathrm{Tr}\left[  \mathrm{ln} (1-eS\slashed{A}) + \sum_{n=1}^{4} \frac{(eS\slashed{A})^n}{n} \right]. 
\ee
The four subtractions in the brackets in (\ref{10}) remove from $\mathrm{det}_5~\Pi_2$ and $\Pi_4$ as well as the tadpole and triangle graphs that are set equal to zero as demanded by $C$-invariance. The remaining $n$-point graphs, $n \geq 5$, contributing to $\mathrm{ln\,det}_5$ can be obtained by expanding $\mathrm{ln\,det}_5$ in powers of $e$.  The gauge invariance of $\mathrm{det}_5$ requires that it depends on $F_{\mu \nu}$ only. 

The representation (\ref{10}) for $\mathrm{det}_5$ is defined only if $S\slashed{A}$ is a compact operator belonging to $\mathscr{S}_r$, $r>4$. The trace ideal $\mathscr{S}_r$ ($1\leq r < \infty$) is defined for those compact operators $T$ with $\mathrm{Tr}(T^{\dagger}T)^{\frac{r}{2}} < \infty$. This means that the eigenstates of $T$ are complete and square-integrable and that the eigenvalues $\lambda_n$ are discrete and satisfy $\sum_n(1/|\lambda_n|^r)<\infty$. General properties of $\mathscr{S}_r$ spaces and the properties of determinants of operators belonging to these spaces may be found in \cite{7}, \cite{8}, \cite{9}, \cite{10}. By a theorem of Seiler and Simon \cite{4}, \cite{5}, \cite{6}, \cite{7}, \cite{11} $S \slashed{A} \in \mathscr{S}_r$, $r>4$ provided $A_{\mu} \in \cap_{r>4}L^{r}(\mathbb{R}^4)$, $m\neq 0$, thereby validating (\ref{10}) for this class of potentials. This restriction on $A_{\mu}$ means that it has no branch points or poles for finite $x$, such as $|x-x_0|^{-\beta}$, $\beta>0$. It also means that $A_{\mu}$ falls off at least as fast as $1/|x|$ for $|x|\to \infty$ and that $A_{\mu}$ is finite at $x=0$.

Since  $S \slashed{A}$ also belongs to $\mathscr{S}_5$, $\mathrm{det}_5$ may be represented as
\be \label{11}
\mathrm{det}_5 = \prod_{n=1}^{\infty} \left[ \left( 1 - \frac{e}{e_n}\right) \exp\left( \sum_{k=1}^{4} \frac{(e/e_n)^k}{k} \right)  \right],
\ee
 where the $\{e_n \}$ are the discrete, complex eigenvalues of $S\slashed{A}$ \cite{8}, \cite{10}. Euclidean $C$-invariance and the reality of $\mathrm{det}_5$ for real $e$ require  that these appear as quartets $\pm e_n$, $\pm \bar{e}_n$ or as complex conjugate pairs. Hence $\mathrm{det}_5$ is an even function of $e$. None of the $e_n$ are on the real axis when $m\neq 0$. Since $\mathrm{det}_5(0)=1$, $\mathrm{det}_5>0$ for real $e\neq 0$. Because $S\slashed{A} \in \mathscr{S}_r$, $r>4$, $\sum_{n}(1/|e_n|)^{4+\epsilon} < \infty$, $\epsilon >0$ so that $\mathrm{det}_5$ is an entire function of $e$ of order 4 \cite{12}. That is, ${\rm det}_5$ is analytic in $e$ in the entire complex $e$-plane with $|{\rm det}_5| < A(\delta) \exp(K(\delta)|e|^{4+\epsilon}) $ for any $\delta>0$ and $A$, $K$ positive constants.
 
 We now relate $\mathrm{det}_5$ to $G$. Since $S\slashed{A} \in \mathscr{S}_{r>4}$, $\mathrm{Tr}(S\slashed{A})^m = \sum_n (1/e_n)^m$ for $m\geq 5$. It is evident from (\ref{11}) that $\mathrm{ln\,det}_5$ has branch points beginning at $|e| = |e_1| \equiv \mathrm{min}\{  |e_n|\}$. Therefore, for $|e|<|e_1| $ the series for $\mathrm{ln\,det}_5$ obtained from (\ref{11}) can be rearranged to give 
 \begin{eqnarray} \nonumber
  \mathrm{ln\,det}_5 & =& -\frac{1}{5} \sum_{n} \left(\frac{e}{e_n}\right)^5 - \frac{1}{6} \sum_{n}  \left(\frac{e}{e_n}\right)^6 - \ldots \\
  			       & =& -\frac{1}{5} \mathrm{Tr}(eS\slashed{A})^5 - \frac{1}{6} \mathrm{Tr}(eS\slashed{A})^6 - \ldots .
 \end{eqnarray}
Within its radius of convergence, $|e|<|e_1|$, this series can be differentiated term-by-term to give
\be \label{13}
e \frac{\partial}{\partial e} \mathrm{ln\,det}_5 = -\mathrm{Tr}[ (eS\slashed{A})^5 + (eS\slashed{A})^6 + \ldots ].
\ee
We can now analytically continue $\mathrm{ln\,\mathrm{det}_5}$  to all $e$ by summing the series:
\be \label{14}
e \frac{\partial}{\partial e} \mathrm{ln\,det}_5 = -e^5 \mathrm{Tr}\left[S\slashed{A}S\slashed{A}S\slashed{A}S\slashed{A}\frac{1}{\slashed{p} - e\slashed{A} + m } \slashed{A} \right].
\ee
Using $G = S + eS\slashed{A}G$ and $\mathrm{Tr}(S\slashed{A})^5=0$ gives the final result
\be \label{15}
e \frac{\partial}{\partial e} \mathrm{ln\,det}_5  = -e^6 \mathrm{Tr}[\slashed{A}S \slashed{A}S \slashed{A}S \slashed{A}S \slashed{A}S\slashed{A}G].
\ee
Thus $G$ is an integral part of $\mathrm{det}_5$, and accordingly $\mathrm{det}_5$ will constrain it.

Let $A_{\mu}$ be scaled by $L$. Suppose for $L\to \infty$ $G(x,y)\neq 0$, except for sets of $x,y$ of measure zero, and finite for $x\neq y$. From (\ref{15}) $| \ln {\rm{det}}_5 (eLF) | = O((eLF)^6)$. If $\ln \rm{det}_5 > 0$ for $L \to \infty$ then such growth on the real $e$-axis is impossible for an entire function of $e$ of order 4 \cite{12}. The only possibility is $\ln \rm det_5 <0$ for  $L \to \infty$. Then the effective action (\ref{9}) would decrease as $\ln {\rm det}_{\text{ren.}} (eLF) \underset{L\to \infty}{\sim} - \Gamma(eL)^6$, where $\Gamma>0$ is a homogeneous function of $F$ of degree 6, thereby establishing the absolute stability of QED. Such an ultrastable QED vacuum is unknown in the literature and contradicts the maximal $O(-(eL)^2\ln(eL))$ decrease of $\ln \mathrm{det}_{\text{ren.}}$ when the $L^2(\mathbb{R}^4)$ zero modes of $\slashed D$ dominate the effective action \cite{Fry}. The calculation in \cite{Fry} does not rely on representation (\ref{9}). Therefore, we conclude for the broad class of potentials for which $\det_5$ is defined that $G$ satisfies for $x\neq y$
\be \label{16}
\lim_{L\to \infty} G(x,y|LA) = 0.
\ee

\noindent Some comments on (\ref{16}): 

\noindent
1. Gauge invariance of (\ref{15}): $G$ in (\ref{15}), (\ref{16}) is gauge-dependent: $G(x,y|A + \partial \lambda) =\exp[ie (\lambda(x)-\lambda(y))] G(x,y|A)$. Thus, it appears that (\ref{15}) cannot be gauge-invariant. However, if $A \to A + \partial \lambda$ in (\ref{15}) then successive iterations of the right-hand side of (15)
  generate gauge-invariant terms so that ${\rm det}_5$ remains gauge invariant. \\


\noindent
2. The conclusion that $G$ vanishes for strong fields is based on $G$ being embedded in ${\rm det}_5$. For potentials with compact support the loop integral defining  ${\rm det}_5$ is cut off at the boundary. So we can only say that (16) holds in the compact support region. \\

\noindent
3. Diamagnetism/paramagnetism: $G$ is constructed  from a complete set of eigenstates of the Hamiltonian $H = (p-A)^2 + \sigma F/2$. It is known that the $O(2) \times O(3)$ symmetric potentials 
$M_{\mu \nu} x_{\nu} a(|x|)$ ($M_{\nu \mu} = - M_{\mu \nu}$, $M$ self- or anti-self dual, $a \sim 1/x^2$, $|x| \to \infty$)  support an unbounded number of $L^2(\mathbb{R}^4)$ zero modes on letting $A_{\mu} \to LA_{\mu}$, $L\to \infty$ \cite{14}. Hence, scaling $A_{\mu}$ does not necessarily enhance the kinetic energy $(p-A)^2$ relative to the spin term $\sigma F/2$. Instead, the spectrum of $H$ can remain at its bottom for arbitrarily large fluctuations of $A_{\mu}$, thereby putting diamagnetism and paramagnetism on an equal footing.  

Random fields can form deep magnetic wells where $F_{\mu \nu}$ is near zero and $A_{\mu}$ is large due to its nonlocality, effectively decoupling the particle's spin from $F_{\mu \nu}$ and enhancing diamagnetism. This and the previous comment on zero modes illustrate the competition between the two terms of $H$ mentioned at the beginning of this paper. \vspace{1em}

\noindent
4. Falloff of $G$: The question arises as to how $G$ decays for potentials on $\mathbb{R}^4$. As stated earlier this is unknown to the author's knowledge. For example, the contribution of the zero modes mentioned above to $G$ with $e=1$ is
\be
\begin{split}
G_{\text{zero modes}}(x,y|LA) = (2\pi^2m)^{-1} \sum_{j=0,1/2,1,\ldots}^J(2j+1) \sum_{m=-j}^j  \\
 \times D_{-jm}^j(\psi, \theta,\varphi) D_{-jm}^{j*}(\psi', \theta',\varphi') (rr')^{2j}R_j(r)R_j(r')\mathbb{K},
\end{split}
\ee
where $\mathbb{K}^T = (0,1,0,0)$, $D_{mm'}^j$ is a spherical harmonic on the four-dimensional sphere, $J$ is the largest value of $j$ for which $2j+2<L$, and 
\be
R_i(r) = \left[ \int_0^{\infty} dr~ r^{4j+3}e^{-2L\int_0^r ds~ sa(s)} \right]^{-1/2} e^{-L \int_0^r ds~s a(s) }.
\ee 
A reliable estimate of $G$'s falloff for $L\to \infty$ remains.  \vspace{0.2em}

Continuing with the potentials introduced above, the scattering states are also known \cite{14}. Their contribution to $G$ requires summing an infinite series of angular momentum states with their varying Clebsch-Gordan coefficients, integrating this sum over energy and taking the $L\to \infty$ limit. As this has not been done yet the falloff of $G$ remains unknown even in this highly symmetric case.  \vspace{1em}


\noindent
5. Growth of random potentials: The potentials on which (\ref{15}) rests may conflict with the growth of a typical $A$ at infinity. By ascribing a property to a typical $A$ we mean all $A$ with the possible exception of a set $\mu(A)$-measure zero (see above Eq. (7)). There is indirect evidence that a typical potential's growth is $|x|^2(\ln |x|)^{\beta}$, $\beta > 1/2$ for $|x| \to \infty$ \cite{15}. The evidence is indirect as the analysis in \cite{15} is for a Gaussian measure whose covariance is that of a free, massive, spin-0 boson. No further work relevant to QED is known to the author. Nevertheless we anticipate that the growth of a typical $A$ will be $|x|^{\alpha}(\ln |x|)^{\beta}$ for some $\alpha, \beta >0$. 

It is generally accepted that the functional integrals for the correlation functions of an interacting field theory have to be calculated in finite volume followed by the removal of the volume cutoffs in the thermodynamic limit. This applies to $\det_5$ and $\det_{\text{ren}.}$ in particular, in which case the growth of a typical potential at infinity will be cut off, allowing $\det_5$ to be defined as above. A possible gauge invariant way to implement the introduction of a volume cutoff in $\ln \det_{\text{ren.}}$ is discussed in Sec. VII of \cite{Fry}. 

The two preceeding paragraphs do not invalidate (\ref{15}) which continues to hold under the assumptions required for its derivation.  \\


In conclusion it has been shown that the strong field behavior of QED's one-loop effective action and the possible decoupling of QED from the remainder of the electroweak model depend on the propagator of a charged fermion in a strong magnetic field. Its strong-field behavior is found to be constrained by the effective action's connection with an entire function of $e$ of order 4 and QED's lack of an ultrastable vacuum, leading to (\ref{16}). 


\end{document}